\documentclass[a4paper,nohyper]{JHEP3}

\newcommand{\dkn}[1]{\ensuremath{\mathrm{d^2}\mathbf{k}_{#1}}}

\newcommand{\qj}[1]{\ensuremath{\mathbf{q}_{#1}}}

\newcommand{\dD}[1]{\ensuremath{\mathrm{d}^{#1}}}
\newcommand{\vq}{\ensuremath{\mathbf{q}}}
\newcommand{\Nc}{\ensuremath{N_c}}
\newcommand{\nf}{\ensuremath{n_f}}
\newcommand{\di}[1]{\ensuremath{\mathrm{d}#1\ }}
\newcommand{\gbmu}{\ensuremath{\bar g_\mu}}
\newcommand{\betao}{\ensuremath{\beta_0}}
\newcommand{\betaonc}{\ensuremath{\frac{\beta_0}\Nc}}
\newcommand{\dkappa}{\ensuremath{\mathrm{d}\kappa}}
\newcommand{\drhof}{\ensuremath{\mathrm{d}\rho_f}}
\newcommand{\vk}[1]{\ensuremath{\mathbf{k}_{#1}}}

\newcommand{\vD}{\ensuremath{\mathbf{\Delta}}}
\newcommand{\Dp}{\ensuremath{\Delta_\perp}}
\newcommand{\knp}[1]{\ensuremath{k_{#1}^+}}

\newcommand{\knm}[1]{\ensuremath{k_{#1}^-}}
\newcommand{\kpn}[1]{\ensuremath{k_{#1\perp}}}
\newcommand{\hs}{\ensuremath{\hat{s}_{12}}}
\newcommand{\qnp}[1]{\ensuremath{q_{#1\perp}}}
\usepackage{amsmath}
\usepackage{cite}
\usepackage{epsfig}
\usepackage{feynmp}
\title{The Quark--Antiquark Contribution to the Fully Exclusive BFKL
  Evolution at NLL Accuracy}
\author{Jeppe R.~Andersen\\Cavendish Laboratory, University of
  Cambridge\\JJ Thomson Avenue\\CB3 0HE\\Cambridge, UK}

\abstract{We calculate the quark--anti-quark contribution to the
  next-to-leading logarithmic corrections to the BFKL kernel, retaining the
  dependence on the momenta of the produced particles. This allows us to
  study the details of the NLL corrections. We demonstrate that the standard
  calculation of the NLL corrections to the scattering of two off-shell
  gluons includes contributions from energies far above that which is probed
  at LL. This explicitly violates energy and momentum conservation in the
  evolution and could be a source of the reported large NLL corrections.  The
  presented calculation is a step towards combining energy and momentum
  conservation with full NLL accuracy in the evolution.}

\preprint{Cavendish-HEP-06/25}
\keywords{QCD, Jets}
\begin{document}
\section{Introduction}
\label{sec:introduction}
The BFKL formalism is currently used both in the description of the small-$x$
evolution of the pdfs, and to obtain the leading logarithmic approximation
(LLA) to the scattering matrix element for e.g.~pure multi-jet and forward
$W$+multi-jet production at hadron colliders. Even though the transverse
scales involved in the two problems are very different, both applications
rely on the same BFKL equation in the description of the evolution of an off-shell
$t$--channel gluon by the emission of several gluons, and
quark--anti-quark pairs at next-to-leading logarithmic accuracy (NLLA). Since
the gluons emitted from the BFKL evolution are relatively hard when the
formalism is applied to the approximation of the hard matrix element, it
proves absolutely crucial to supplement the solution of the BFKL equation
with energy and momentum conservation. While this is possible using the LL
BFKL kernel, the derivation of the NLL BFKL kernel already contains phase
space integrals over the invariant mass of two emitted gluons and a
quark--anti-quark pair. In the standard
analysis\cite{Fadin:1998py,Ciafaloni:1998gs}, the NLL corrections are found
to be size-able, and it therefore seems imperative to include them in the
calculation of any quantity which is to be confronted with reality.

In this paper we will recalculate the quark-contribution to the NLL BFKL
kernel in a form that will allow us to combine the evolution to full NLL
accuracy with energy and momentum conservation\cite{Andersen:2006sp}. We will
use the obtained result to study the make-up of the NLL corrections.
Specifically, we will show that in the standard analysis, due to the lack of
energy and momentum conservation, the NLL corrections to a Reggeon-Reggeon
scattering at a given energy receives contributions from significantly larger
energies. This could be a significant and unintended source of the observed
large NLL corrections.

In the interest of clarity and conciseness in our argumentation, we have
chosen to include all intermediate results in the derivation of the NLL BFKL
kernel.

\section{Exclusive States to Next-to-Leading Logarithmic Accuracy}
\label{sec:excl-stat-next}
We will begin this section by recalling the results of the calculation of the
NLL corrections to the BFKL kernel. In the following sections we will
recalculate the quark--anti-quark contribution using a phase space slicing
method, which allows us to study directly the energy dependence of the
corrections to the Reggeon-Reggeon collision.

The BFKL equation governing the evolution in rapidity $y$ of an off-shell
gluon exchanged in the $t$-channel can be written as
\begin{align}
  \label{eq:BFKLdiffEqn}
  \frac{\partial}{\partial y}\ f(\qj a,\qj b,y)=\int\!\dD {D-2} \vq\ K(\qj
  a,\vq)\ f(\vq,\qj b,y), 
\end{align}
where $f(\qj a,\qj b,y)$ is the gluon Green's function, and we have used
boldface to denote transverse vectors. The BFKL kernel $K(\qj 1,\qj 2)$ is
split into contributions from virtual and real corrections, embedded in the
trajectory $\omega(\qj 1^2)$ and real emission kernel $K_r(\qj 1,\qj 2)$,
respectively
\begin{align}
  \label{eq:Kernel}
  K(\qj 1,\qj 2)=2\ \omega(\qj{1}^2)\ \delta^{D-2}(\qj 1-\qj 2)+K_r(\qj 1,\qj 2).
\end{align}
The first two terms in the perturbative expansion of each of these terms are known
\begin{align}
  \label{eq:TrajectoryKernelExpansion}
  \omega(\vq^2)=\omega^{(1)}(\vq^2)+\omega^{(2)}(\vq^2)+\cdots,\quad K_r(\qj
  1,\qj 2)=K_r^{(1)}(\qj 1,\qj 2)+K_r^{(2)}(\qj 1,\qj 2)+\cdots.
\end{align}
The leading contributions are given by
\begin{align}
  \label{eq:LLReggeTrajectoryandKernel}
  \omega^{(1)}(\vq^2)=-\bar{g}_\mu^2\left(\frac 2 \varepsilon+
    2\ln\frac{\vq^2}{\mu^2}\right), \quad K_r^{(1)}(\qj 1,\qj 2)=\frac{4\ \bar
  g_\mu^2\ \mu^{-2\varepsilon}}{\pi^{1+\varepsilon}\ \Gamma(1-\varepsilon)}\frac
1 {(\qj 1 - \qj 2)^2}
\end{align}
with
\begin{align}
  \label{eq:coupling}
  \bar g_\mu^2=\frac{g_\mu^2\ \Nc\ \Gamma(1-\varepsilon)}{(4\pi)^{2+\varepsilon}}
\end{align}
The next-to-leading logarithmic correction to the trajectory is\cite{Fadin:1998py}
\begin{align}
  \begin{split}
    \label{eq:NLLReggeTrajectory}
    \omega^{(2)}(\vq^2)=-\gbmu^4\bigg[\betaonc\left(\frac 1 {\varepsilon^2}
      -\ln^2\left(\frac{\vq^2}{\mu^2}\right)\right) + \left(\frac{67}9-
      \frac{\pi^2}3-\frac {10}9\frac\nf\Nc\right)\left(\frac 1 \varepsilon +
      2\ln\left(\frac{\vq^2}{\mu^2}\right)\right)\\
    -\frac{404}{27}+2\zeta(3)+\frac{56}{27}\frac{\nf}{\Nc}\bigg]
  \end{split}
\end{align}
and to the real emission kernel\cite{Fadin:1998py}
\begin{align}
  \begin{split}
    \label{eq:NLLRealEmission}
    K_r^{(2)}(\qj 1,\qj 2)=\frac{4\ \gbmu^4\
      \mu^{-2\varepsilon}}{\pi^{1+\varepsilon}\
      \Gamma(1-\varepsilon)}\Bigg\{\frac 1 {(\qj 1 - \qj
      2)^2}\bigg[\betaonc\frac 1\varepsilon \left(1-\left(\frac{(\qj 1 - \qj
          2)^2}{\mu^2}\right)^\varepsilon
      (1-\varepsilon^2 \frac{\pi^2}6\right)\\
    +\left(\frac{(\qj1-\qj2)^2}{\mu^2}\right)^\varepsilon\left(\frac{67}9
      -\frac{\pi^2}3 -\frac{10}9\frac\nf\Nc + \varepsilon
      \left(-\frac{404}{27} + 14\zeta(3)
        +\frac{56}{27}\frac\nf\Nc\right)\right)\bigg]\\
    -\left(1+\frac\nf{\Nc^3}\right)\frac{2\qj1^2\qj2^2-3(\qj1.\qj2)^2}{16\qj1^2\qj2^2}\left(
      \frac 2 {\qj2^2}+\frac 2 {\qj1^2}+\left(\frac 1 {\qj2^2}-\frac 1
        {\qj1^2}\right)\ln\frac{\qj1^2}{\qj2^2}\right)-\frac 1
    {(\qj1-\qj2)^2}\left(\ln\frac{\qj1^2}{\qj2^2}\right)^2\\
    +\frac{2(\qj1^2-\qj2^2)}{(\qj1-\qj2)^2(\qj1+\qj2)^2}\left(\frac 1 2
      \ln\left(\frac{\qj1^2} {\qj2^2}\right) \ln\left(\frac{\qj1^2\qj2^2(\qj1
          -\qj2)^4}{(\qj1^2+\qj2^2)^4
        }\right)+L\!\left(-\frac{\qj1^2}{\qj2^2}\right)-L\!\left(-\frac{\qj2^2}{\qj1^2}\right) \right)\\
    -\left(3+\left(1+\frac\nf{\Nc^3}\right)\left(1-\frac{(\qj1^2+\qj2^2)^2}{8\qj1^2\qj2^2} - \frac{2\qj1^2\qj2^2-3\qj1^4-3\qj2^4}{ 16\qj1^4\qj2^4}(\qj1.\qj2)^2 \right)\right)\int_0^\infty\!\!\!\di x\frac{\ln\left|\frac{1+x}{1-x}\right|}{\qj1^2+x^2\qj2^2}\\
-\left(1-\frac{(\qj1^2-\qj2^2)^2}{(\qj1-\qj2)^2(\qj1+\qj2)^2}\right)\left(\int_0^1\di
z-\int_1^\infty\di z\right)\frac{\ln\frac{(z\qj1)^2}{\qj2^2}}{(\qj2-z\qj1)^2}\Bigg\},
  \end{split}
\end{align}
where 
\begin{align}
  \label{eq:dilog}
  L(z)=\int_0^z\frac{\di t}t\ln(1-t),\quad \zeta(n)=\sum_{k=1}^\infty
  k^{-n},\quad\beta_0=\left(\frac{11}3\Nc-\frac 2 3 \nf\right).
\end{align}

The contribution to the real emission kernel at NLL accuracy has three
sources: 1) the one-loop corrections to one gluon emission in multi Regge
kinematics, 2) two-gluon emission in quasi multi Regge kinematics, and 3)
quark--anti-quark production in quasi multi Regge kinematics:
\begin{align}
  \label{eq:KernelSplit}
    K_r^{(2)}(\qj 1,\qj 2)=K_r^{(2),g}(\qj 1,\qj 2)+K_r^{(2),gg}(\qj 1,\qj
    2)+K_r^{(2),q\bar q}(\qj 1,\qj 2)
\end{align}
It is the purpose of this paper to calculate the contribution $K_r^{(2),q\bar
  q}(\qj 1,\qj 2)$ in a form that retains not only the dependence on the sum
of the transverse momenta of the emitted quark pair $(\qj 1-\qj 2)$, but the
dependence on the full momenta of both emitted particles.

The divergences of $K_r^{(2),q\bar q}(\qj 1,\qj 2)$ will fall into two
categories; one at ${\boldmath \Delta}^2=(\qj 1-\qj 2)^2\to0$ regularised by
the quark contribution to the trajectory at NLL, and an explicit pole for
$\epsilon\to 0$, regularised by the quark contribution to $K_r^{(2),g}(\qj
1,\qj 2)$. The one-gluon production vertex for the collision of two Reggeised
gluons at NLL accuracy can be calculated from a tree approximation for the
quasi-multi-Regge kinematics combined with $t$--channel unitarity
relations\cite{Fadin:1993wh,Fadin:1993qb,Fadin:1996yv}, or extracted directly
from the one loop 5 gluon amplitude in the helicity
basis\cite{DelDuca:1998cx}. Expanded in $D=4+2\varepsilon$ it is given
by\cite{Fadin:1998sh}:
\begin{align}
  \begin{split}
    \label{eq:onegluonvertexNLL}
    &K_r^{(2),g}(\qj 1,\qj 2)=\frac{\gbmu^2\mu^{-2\varepsilon}}{\pi^{1+\varepsilon} \Gamma(1-\varepsilon)} \frac 4 {\vk1^2}\Bigg(1+\gbmu^2\Bigg[-2\left(\frac{\vk1^2}{\mu^2}\right)^\varepsilon \left( \frac{1}{\varepsilon^2}-\frac{\pi^2}2+2\varepsilon\zeta(3)\right)+\frac \betao \Nc\frac 1 \varepsilon\\
    &+\frac{3\vk1^2}{\vq_1^2-\vq_2^2}\ln\left(\frac{\vq_1^2}{\vq_2^2}\right)-\ln^2\left(\frac {\vq_1^2}{\vq_2^2}\right)+\left(1-\frac{\nf}{\Nc}\right)\left(\frac{\vk1^2}{\vq_1^2-\vq_2^2}\left(1-\frac{\vk1^2(\vq_1^2+\vq_2^2+4\vq_1.\vq_2)}{3(\vq_1^2-\vq_2^2)^2} \right)\ln\left(\frac{\vq_1^2}{\vq_2^2}\right)\right)\\
    &-\frac{\vk1^2}{6\vq_1^2\vq_2^2}(\vq_1^2+\vq_2^2+2\vq_1.\vq_2)+\frac{(\vk1^2)^2(\vq_1^2+\vq_2^2)}{6\vq_1^2\vq_2^2(\vq_1^2-\vq_2^2)^2}(\vq_1^2+\vq_2^2+4\vq_1.\vq_2)\Bigg]\Bigg)
  \end{split}
\end{align}
The divergent piece from the quark contribution is simply
\begin{align}
  \label{eq:onegluonvertexNLLnfdivergence}
  \frac{\gbmu^4\mu^{-2\varepsilon}}{\pi^{1+\varepsilon}
    \Gamma(1-\varepsilon)} \frac 4 {\vD^2}
  \frac{-2}3\frac{\nf}{\Nc}\frac 1 \varepsilon.
\end{align}

\subsection{The Quark-antiquark Contribution to the NLL Corrections}
\label{sec:quark-antiq-contr}
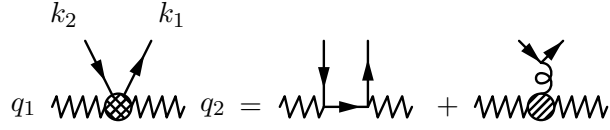
\begin{figure}
  \centering
  \begin{align*}
    \parbox{20mm}{\begin{fmffile}{NLLquarkantiquark2}
        \begin{fmfgraph*}(50,50)
          \fmfset{arrow_len}{3mm} \fmfset{arrow_ang}{15} \fmfpen{1pt}    
          \fmfstraight \fmfleft{i0} \fmfright{o0}
          \fmftop{t0,t1,t2,t3,t4}
          \fmf{phantom,tension=1}{i0,v1,o0} 
          \fmffreeze 
          \fmf{zigzag}{i0,v1,o0}
          \fmffreeze
          \fmf{quark}{t1,v1,t3}
          \fmfv{decor.shape=circle,decor.filled=hatched,decor.size=5thick}{v1}
        \fmflabel{$q_1$}{i0}
        \fmflabel{$k_2$}{t1}
        \fmflabel{$k_1$}{t3}
        \fmflabel{$q_2$}{o0}
        \end{fmfgraph*}
      \end{fmffile}}
\quad =\
    \parbox{20mm}{\begin{fmffile}{NLLquarkantiquark2a}
      \begin{fmfgraph*}(50,50)
        \fmfset{arrow_len}{3mm} \fmfset{arrow_ang}{15} \fmfpen{1pt}    
        \fmfstraight \fmfleft{i0} \fmfright{o0}
        \fmftop{t0,t1,t2,t3}
        \fmf{phantom,tension=1}{i0,v1,v2,o0} 
        \fmffreeze 
        \fmf{zigzag}{i0,v1}
        \fmf{zigzag}{v2,o0}
        \fmf{quark}{t1,v1,v2,t2}
      \end{fmfgraph*}
    \end{fmffile}}+\ 
    \parbox{20mm}{\begin{fmffile}{NLLquarkantiquark2b}
      \begin{fmfgraph*}(50,50)
        \fmfset{arrow_len}{3mm} \fmfset{arrow_ang}{15} \fmfpen{1pt}    
        \fmfstraight \fmfleft{i0} \fmfright{o0}
        \fmftop{t0,t1,t2,t3}
        \fmf{phantom,tension=1}{i0,v1,o0} 
        \fmffreeze 
        \fmf{zigzag}{i0,v1,o0}
        \fmf{gluon}{v1,v2}
        \fmf{quark}{t1,v2,t2}
        \fmfv{decor.shape=circle,decor.filled=shaded,decor.size=5thick}{v1}
      \end{fmfgraph*}
    \end{fmffile}}
  \end{align*}
  \caption{The quark--anti-quark production amplitude in Reggeon-Reggeon
    collisions (LHS) can be obtained as the sum of the contributing diagrams (RHS)
    according to the effective Feynman rules of
    Ref.\cite{Lipatov:1995pn}. The zig-zag lines denote the Reggeons
    (off-shell gluons). The BFKL evolution consists of a string of such
    vertices (and their LL and NLL gluon emission counterparts) connected
    with Reggeised gluon propagators.}
  \label{fig:quarkcontr}
\end{figure}

The Reggeon-Reggeon-$q\bar q$-vertex can be calculated using the Feynman
rules of the effective action in the high-energy limit in
Ref.\cite{Lipatov:1995pn}. The contributing diagrams are shown in
Fig.~\ref{fig:quarkcontr}. The benefit of this framework of obtaining the
scattering amplitudes is that it gives an immediate picture of the
contributing sub-processes. However, the RR$q\bar q$ (or $g^* g^*q\bar
q$) vertex can also be extracted directly from the full tree-level $gg\to
ggq\bar q$ scattering amplitude in the helicity basis\cite{DelDuca:1996me}.
In terms of the operators of the effective action of
Ref.\cite{Lipatov:1995pn}, the result for the $g^*g^*q\bar q$ vertex
$\gamma_{i_1i_2}^{Q\bar Q}(q_1,q_2,k_1,k_2)$ is
\begin{align}
  \label{eq:gammaQQbar}
  \gamma_{i_1i_2}^{Q\bar Q}(q_1,q_2)=\frac 1 2 g^2\bar
  u(k_1)\big[t^{i_1}t^{i_2}b(k_1,k_2)-t^{i_2}t^{i_1}\overline{b(k_2,k_1)} \big]v(k_2)
\end{align}
where $t^{i_1},t^{i_2}$ are the colour group generators for the fundamental representation.
In terms of the
helicity amplitude $A^{\bar q q}_{+-}$ the spin, colour and flavour sum is
\begin{align}
  \begin{split}
    &\sum_{i_1,i_2,f}\left|\gamma_{i_1i_2}^{Q\bar
        Q}(q_1,q_2,k_1,k_2)\right|^2\\
    &=\frac{2\ g^4 (\Nc^2-1)\nf}{\Nc}\Bigg[(\Nc^2-1)\left(\left|A^{\bar q
          q}_{+-}(k_1,k_2)\right|^2+\left|A^{\bar q
          q}_{+-}(k_2,k_1)\right|^2\right)+2\ \mathrm{Re}\ A^{\bar q
          q}_{+-}(k_1,k_2) A^{\bar q
          q}_{+-}(k_2,k_1)\Bigg],
  \end{split}
\end{align}
and in terms of the light-cone momenta and complex transverse momenta the
amplitude $A^{\bar q q}_{+-}$ is given by\cite{DelDuca:1996me}
\begin{align}
  \begin{split}
    \label{eq:Aqbarqplusminus}
    A^{\bar q
      q}_{+-}(k_1,k_2)=&-\sqrt{\frac{\knp1}{\knp2}}\Bigg\{\frac{\knp2|\qnp2|^2}{(\knp1+\knp2)\hs}+\frac{\knm2\kpn2|\qnp1|^2} {\kpn1(\knm1+\knm2)\hs}+ \frac{\knp2\kpn1^*(\qnp2+\kpn2)}{\knp1\hat t}\\
    &+\frac{(\qnp2+\kpn2)[\knm1\knp2-\kpn1^*\kpn2-(\qnp2^*+\kpn2^*)\kpn2]}{\kpn1\hs} - \frac{|\kpn2|^2}{\hs} \Bigg\}
  \end{split},
\end{align}
where $\hat s_{12}=(k_1+k_2)^2$ and $\hat t=(k_2+q_2)^2=(q_1-k_1)^2$.


The quark contribution to the NLL corrections to the BFKL kernel,
$K_r^{(2),q\bar q}(\qj 1,\qj 2)$ in Eq.~\eqref{eq:qqbarintegrated}, is in the
traditional analysis\cite{Fadin:1998py,Ciafaloni:1998gs} obtained as the
integral over the phase space of the emitted quark-antiquark pair of the
fully exclusive $RRq\bar q$-vertex
\begin{align}
  \label{eq:qqbarexlusiveintegrated}
  K_r^{(2),q\bar q}(\qj 1,\qj 2)=\frac 1 {2\qj1^2\qj2^2}\frac 1
  {\Nc^2-1}\int\dkappa\ \drhof\ \delta^{(D)}(q_1-q_2-k_1-k_2)\ \sum_{i_1,i_2,f}\left|\gamma_{i_1i_2}^{Q\bar Q}(q_1,q_2,k_1,k_2)\right|^2
\end{align}
where $\kappa=(q_1-q_2)^2=(k_1+k_2)^2=\hat s_{12}$ is the invariant mass, and
\begin{align}
  \label{eq:drhof}
  \drhof=\prod_{n=1,2}\frac{\dD{D-1} k_n}{(2\pi)^{D-1} 2E_n}.
\end{align}
The result of the integration in $D=4+2\epsilon$ dimensions can be written as\cite{Fadin:1997hr}
\begin{align}
  \begin{split}
    \label{eq:qqbarintegrated}
    K_r^{(2),q\bar q}(\qj 1,\qj 2)=\frac{4\ \gbmu^4\ \mu^{-2\varepsilon}\
      \nf} {\pi^{1+\varepsilon}\ \Gamma(1-\varepsilon)\ \Nc^3} \Bigg\{\Nc^2
    \Bigg[\frac 1 {(\qj1-\qj2)^2}\left(\frac{(\qj1-\qj2)^2}
      {\mu^2}\right)^\varepsilon\frac 2 3\left( \frac 1 \varepsilon - \frac 5
      3 +\varepsilon\left(\frac{28}9-\frac{\pi^2}6\right)\right) \\
    +\frac 1
    {\qj1^2-\qj2^2}\left(1-\frac{(\qj1-\qj2)^2(\qj1^2+\qj2^2+4\qj1.\qj2)}{3(\qj1^2-\qj2^2)^2}\right)\ln\left(\frac{\qj1^2}{\qj2^2}\right)\\
+\frac{(\qj1-\qj2)^2}{(\qj1^2-\qj2^2)^2}\left(2-\frac{(\qj1-\qj2)^2(\qj1^2+\qj2^2)}
{3\qj1^2\qj2^2}\right)+\frac{(2(\qj1-\qj2)^2-\qj1^2-\qj2^2)}{3\qj1^2\qj2^2}
\Bigg]\\
-\Bigg[1-\frac{(\qj1^2+\qj2^2)^2}{8\qj1^2\qj2^2} - \frac{2\qj1^2\qj2^2-3\qj1^4-3\qj2^4}{ 16\qj1^4\qj2^4}(\qj1.\qj2)^2
\Bigg]\int_0^\infty\!\!\!\di
x\frac{\ln\left|\frac{1+x}{1-x}\right|}{\qj1^2+x^2\qj2^2}\\
+\frac{3(\qj1.\qj2)^2-2\qj1^2\qj2^2}{16\qj1^2\qj2^2}\left(
      \frac 2 {\qj2^2}+\frac 2 {\qj1^2}+\left(\frac 1 {\qj2^2}-\frac 1
        {\qj1^2}\right)\ln\frac{\qj1^2}{\qj2^2}\right)
    \Bigg\}
  \end{split}
\end{align}
This result enters directly in the real emission kernel of
Eq.~\eqref{eq:NLLRealEmission}, and it is the validity of this result we will
discuss. The problem we wish to highlight is the r\^ole of the
$\delta$--functional in the integrations of
Eq.~\eqref{eq:qqbarexlusiveintegrated}. While each amplitude
(e.g.~$\gamma_{i_1i_2}^{Q\bar Q}(q_1,q_2,k_1,k_2)$ and the gluonic
counterparts) conserves energy and momentum (i.e.~all 4 components of the
momentum vectors), the conservation of the longitudinal and energy component
is lost in
Eqs.~\eqref{eq:qqbarexlusiveintegrated},\eqref{eq:qqbarintegrated}, in order
to write the result in a form that depends only on transverse vectors. In
Eq.~\eqref{eq:qqbarintegrated} there is no relation between the longitudinal
and energy component of the reggeon momenta, and the momenta of the
quark--anti-quark pair. This has two effects.  First, energy and momentum
conservation is broken even at NLL accuracy when connecting the vertices to
form the BFKL evolution (this effect was already highlighted in
Ref.\cite{Andersen:2006sp}). This part of the problem can be solved by using
the framework of the direct solution of the BFKL
evolution\cite{Andersen:2003an,Andersen:2003wy,Andersen:2006sp}. Secondly,
within each Lipatov vertex the NLL corrections to the reggeon-reggeon
scattering are not constrained to the correct energy probed in the
corresponding LL process, but rather integrated over all energies. This
obviously not only violates energy and momentum conservation at the level of
each vertex, but also exaggerates the contribution of the NLL corrections.
This problem, and its solution, is the focus of the remaining study.

We will calculate the quark contribution to the NLL corrections by performing
the integral in Eq.~\eqref{eq:qqbarexlusiveintegrated}, but regularise the
result by the phase space slicing method, and combine it with the pole in
$\varepsilon$ of Eq.~\eqref{eq:onegluonvertexNLLnfdivergence} from the quark
contribution to the loop corrections of the one-gluon production Lipatov
vertex. The result will be finite for ${\boldmath \Delta}^2\not=0$. In
performing the integration, we will parametrise the phase space for fixed
${\boldmath\Delta}$ in terms of the transverse momenta of the anti-quark
(\kpn 1) and half the rapidity separation between the quark and the
antiquark, $\eta$ (the quark and anti-quark are produced with rapidity $\eta$
and $-\eta$). We find
\begin{align}
  \label{eq:PhaseSpace}
  \int\dkappa\ \drhof\ \delta^{(D)}(q_1-q_2-k_1-k_2)=\int\mathrm{d}\eta\int\frac{\dD{D-2}k_1}{(2\pi)^{2(D-1)}},
\end{align}
which once again illustrates that the NLL corrections to the Lipatov vertex
includes contribution from a range of energies and not just that probed by
the LL vertex (i.e.~for a given $\qj1,\qj2$ the NLL corrections include
contributions from all invariant masses and energies in the Reggeon-Reggeon
collision, not just the one probed at LL). The parametrisation of the phase space
in Eq.~\eqref{eq:PhaseSpace} allows for an immediate implementation of the
integration in Eq.~\eqref{eq:qqbarexlusiveintegrated}.

\begin{figure}
  \centering
  \epsfig{file=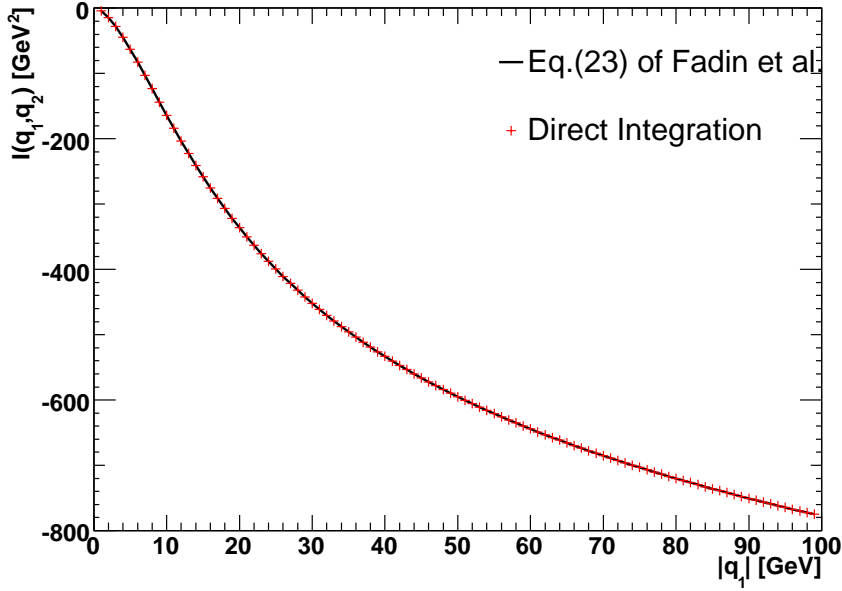,width=.8\textwidth}
  \caption{The quantity $I(\qj 1,\qj 2)$ of Eq.~\eqref{eq:finitepart} for
    $\qj2=(0,15)$~GeV and $\qj1$ varying along $(1,0)$ compared between the
    results obtained in this paper and the corrected results of
    Ref.\cite{Fadin:1997hr}.}
  \label{fig:IFcomp}
\end{figure}

Symmetry properties in the divergent part of the amplitude ensures that the
integral over the $1/\Nc^2$ suppressed contribution of the quark--anti-quark
vertex is finite, and the result for
\begin{align}
    \label{eq:finitepart}
    &I(\qj 1,\qj 2)=\\
    &64\ (2\pi)^3 \int\mathrm{d}\eta\int\frac{\dkn 1}{(2\pi)^{6}} \Bigg[2\
    \mathrm{Re}\ A^{\bar q q}_{+-}(k_1,k_2) A^{\bar q q}_{+-}(k_2,k_1) -
    \left (\left|A^{\bar q q}_{+-}(k_1,k_2)\right|^2+\left|A^{\bar q
          q}_{+-}(k_2,k_1)\right|^2\right)\Bigg]\nonumber
\end{align}
can be directly compared to Eq.~(23) of Ref.\cite{Fadin:1997hr}, which up to
constants form the $1/\Nc^2$--suppressed part of
Eq.~\eqref{eq:qqbarintegrated}. The result is shown in Fig.~\ref{fig:IFcomp}
for fixed $\qj2=(0,15)$~GeV as a function of $\qj1=(x,0)$~GeV. We find
complete agreement, after correcting a misprint in Eq.~(23) of
Ref.\cite{Fadin:1997hr} (the denominator $216\ \qj1^2\ \qj2^2$ should read
$16\ \qj1^2\ \qj2^2$).
The complete agreement verifies that our implementation of the amplitudes and
parametrisation of phase space is correct.

\subsection{Regularisation}
\label{sec:regularisation}
We now want to regularise the integral in
Eq.~\eqref{eq:qqbarexlusiveintegrated} in order to implement the quark
contribution to the fully exclusive Lipatov vertex at NLL. We will start by
rewriting\cite{Fadin:1996nw,DelDuca:1996me} the amplitude $A^{\bar q
  q}_{+-}(k_1,k_2)$
\begin{align}
  \begin{split}
    \label{eq:AFaL}
    A^{\bar q q}_{+-}(k_1,k_2) =&
    -\sqrt{\frac{1-x}{x}}\Bigg[\frac{\kpn1^*(\qnp2+\kpn2)}{\hat
      t}+\frac{x|\qnp1|^2\kpn1^*\kpn2}{|\Dp|^2(|\kpn1-x\Dp|^2+x(1-x)|\Dp|^2)}\\
    &-\frac{x(1-x)\qnp1\qnp2^*}{\Dp^*(\kpn1-x\Dp)}+\frac{x
      \qnp1^*\qnp2\kpn1^*}{|\Dp|^2(\kpn1^*-x\Dp^*)}+\frac{x\qnp2^*}{\Dp^*}\Bigg]
  \end{split}
\end{align}
with 
\begin{align}
  \begin{split}
    \label{eq:that}
    \hat t=&-\frac{|\kpn1-x\qnp1|^2+x(1-x)|\qnp1|^2}{x}, \qquad
    x=\frac{\knp1}{\knp1+\knp2}\\
    \hat s_{12}&=\frac{\left|\kpn1-x\Dp \right|}{x(1-x)}, \qquad \Dp=\kpn1+\kpn2
  \end{split}
\end{align}
The quark--anti-quark contribution has two divergences: A singularity at
$\knp1-x\Dp\to0$ and one at $\Dp^2=0$. When constructing the regularised
vertex, the first type of singularity is regularised by the corresponding
singularity from the one-loop corrections to the one-gluon production,
whereas the second type of singularity is regularised by the corresponding
one in the gluon trajectory. In the framework of the direct solution to the
BFKL evolution, the regularisation of the latter divergence was
performed already in Ref.\cite{Andersen:2003an,Andersen:2003wy}. In the present
context we therefore need only to consider $\Dp^2\not=0$.  In order to implement
the exclusive NLL Lipatov vertex, we would need to perform the regularisation
of the divergence at $\kpn1-x\Dp\to0$ using a phase space slicing method.
Technically, we will perform the phase space integration by Monte Carlo
techniques. If $A^{\bar q q}_{+-}(k_1,k_2)$ is probed for
$\kpn1-x\Dp>\lambda$ with $\lambda$ small (our reported results are stable
for variations of $\lambda$ between $10^{-2}$~GeV and $10^{-10}$~GeV), the
standard result of Eq.~\eqref{eq:Aqbarqplusminus} is used. When
$\kpn1-x\Dp\le\lambda$, in addition to the non-divergent pieces we return the
average value of the integral of $|A^{\bar q q}_{+-,\mathrm{div}}(k_1,k_2)|^2$ combined
with the appropriate term from the one--loop quark contribution to the
one-gluon emission vertex, evaluated for $\varepsilon\to0$.

It turns out that only the square of the two terms in Eq.~\eqref{eq:AFaL}
that have an explicit divergence at $\kpn1\to x\Dp$ lead to a pole in
$\varepsilon$. All cross terms integrate to zero when integrated for
$\kpn1-x\Dp\le\lambda$.  We therefore find
\begin{align}
  \begin{split}
    \label{eq:divsquare}
    |A^{\bar q
      q}_{+-,\mathrm{div}}(k_1,k_2)|^2&\to\frac{1-x}{x}\left(\left|\frac{x(1-x)\qnp1\qnp2^*}{\Dp^*(\kpn1-x\Dp)}\right|^2+\left|\frac{x
          \qnp1^*\qnp2\kpn1^*}{|\Dp|^2(\kpn1^*-x\Dp^*)}\right|^2\right)\\
    &=\frac{1-x}{x}\frac{\vq_1^2\vq_2^2}{\mathbf{\Delta}^2(\vk1-x \mathbf{\Delta})^2}\left(x^2(1-x)^2+x^4\right).
  \end{split}
\end{align}
Using
\begin{align}
  \label{eq:detadx}
  \mathrm{d}\eta=\frac{\mathrm{d}x}{2x(1-x)}
\end{align}
and the substitution $\vk1-x\mathbf{\Delta}\to\mathbf{l}$ together with the following
result
\begin{align}
  \label{eq:phasespaceint1}
  \int_{|l|=0}^{|l|=\lambda}\frac{\mathrm{d}^{D-2}\mathbf{l}}{l^2}=\frac{\pi^{1+\varepsilon}}{\Gamma(1+\varepsilon)}\frac{1}{\varepsilon} (\lambda^2)^\varepsilon
\end{align}
with $D=4+2\varepsilon$, we can obtain the Laurent series in $\varepsilon$
for the relevant phase space integral of each of the contributing terms. The result is
\begin{align}
  \frac{g^4\mu^{2\varepsilon}}{\mathbf{\Delta}^2\
    (2\pi)^6}\nf\Nc\frac{\pi^{1+\varepsilon}}{3\ \Gamma(1+\varepsilon)}\frac 1 \varepsilon (\lambda^2)^\varepsilon
\end{align}
The generated divergence cancels with the divergent part of the one--loop quark
contribution to the one-gluon emission vertex in
Eq.~\eqref{eq:onegluonvertexNLLnfdivergence}. 

\section{Discussion}
\label{sec:discussion}
\begin{figure}
  \centering
  \epsfig{file=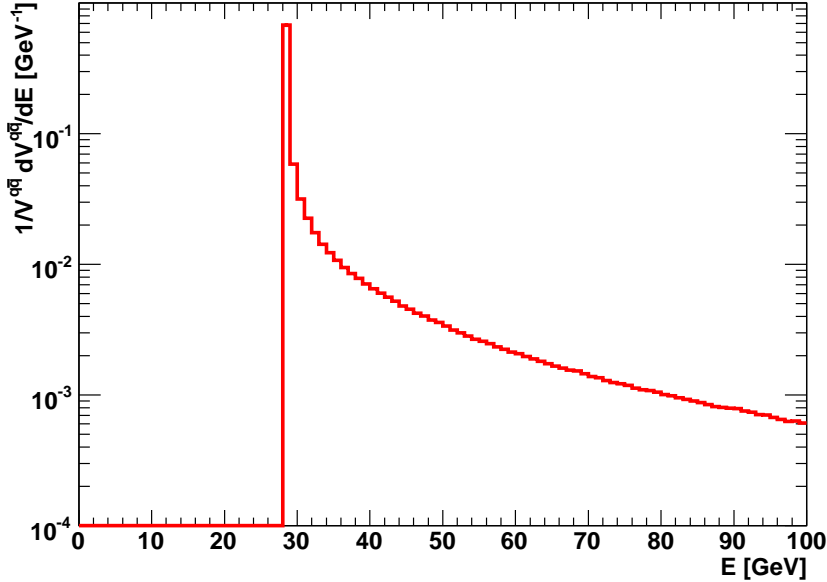,width=.8\textwidth}
  \caption{The energy distribution of the contribution to the NLL corrections
  to a Reggeon-Reggeon collision of $\sqrt 2 \cdot 20$~GeV in the standard approach.}
  \label{fig:edist}
\end{figure}

Having thus constructed the regularised quark contribution, we can calculate
the contribution to the BFKL kernel as in
Eq.~\eqref{eq:qqbarexlusiveintegrated} (remembering that also the divergent
contribution from the virtual corrections are taken into account). The Monte
Carlo method allow us to study directly the differential distribution in any
variable. Of particular interest is the distribution in the energy of the
Reggeon-Reggeon collision. The LL contribution come from the emission of a
single gluon of energy $|\qj1-\qj2|$. However, when the NLL corrections are
calculated as in Eq.~\eqref{eq:qqbarexlusiveintegrated}, contributions arise
from all energies above this scale. In Fig.~\ref{fig:edist} we have plotted
the distribution in the energy of the Reggeon-Reggeon collisions contributing
to the NLL corrections to the LL result. We have chosen $\qj1$ and $\qj2$ to
be perpendicular, with $|\qj1|=|\qj2|=20$~GeV, however the findings have
general validity. The average energy of the Reggeon-Reggeon collisions
included in the NLL corrections calculated according to
Eq.~\eqref{eq:qqbarexlusiveintegrated} is $\approx40\%$ larger than the
energy in the LL setup (for all angles and for an arbitrary value of
$|\qj1|=|\qj2|$). Furthermore, the average rapidity separation between the
quark and the anti-quark is $\approx .56$ units of rapidity. Often, the
contribution supposed to described the radiative corrections to a single
gluon emission would be described as a two-jet configuration! All the
quark--anti-quark configurations included in the integrations of
Eq.~\eqref{eq:qqbarexlusiveintegrated} must be included in the evolution to
obtain overall NLL accuracy. But assigning them all to a single transverse
scale will potentially greatly exaggerate the impact of these NLL corrections
on the evolution\footnote{A similar problem will appear in the calculation of
  the impact factors}.

However, this is a problem that cannot be resolved as long as only transverse
degrees of freedom are considered in the governing BFKL equation. We
therefore propose a framework which will perform the BFKL evolution fully
differentially in the momenta of all emitted particles, i.e.~not only using
the fully differential BFKL
equation\cite{Andersen:2006sp,Andersen:2003wy,Andersen:2003an}
\begin{align}
\label{eq:BFKLdiffdiffeqn}
  \frac{\partial}{\partial y_j}\
  \frac{\partial}{\partial \kpn j}\ f(\qj a,\{ \kpn i, y_i\},\qj b,y)= K\left(\qj
  a+\sum_{i=1}^{j-1}\kpn i,\qj
  a+\sum_{i=1}^{j}\kpn i \right)\ f(\qj a,\{ \kpn i, y_i\},\qj b,y), 
\end{align}
but also keeping the NLL corrections fully exclusive as detailed in this
study for the quark contributions. This will ensure not only the conservation
of all components of the momentum vector, but also that for a given set of
Reggeon momenta, only the relevant NLL corrections are included. The result
of such evolution would be in complete agreement with the description of the
Reggeisation of $t$-channel gluon amplitudes according to
Ref.\cite{Fadin:2006bj}, when also here the NLL vertices are kept exclusive.
Of course the result of the standard BFKL formalism can be obtained by
suitable phase space integrations and ignorance of the conservation of
longitudinal momentum.

The large spread in energies of the contributions to the NLL corrections will
have sizeable effects in a realistic application of the BFKL framework. In
the application of the BFKL framework to the description of the small-$x$
behaviour of partons, the calculated NLL corrections to the kernel cannot be
ascribed to a single value of $x$. And in the emerging framework of applying
the BFKL evolution in the description of the production of multiple hard
jets, the NLL corrections would be weighted with different pdf-factors,
according to their relevant contribution to the centre of mass energy. This
effect will change the impact of the NLL corrections, and could further
reduce their importance.

\section{Conclusions}
\label{sec:conclusions}
We have presented a new calculation of the quark contribution to the NLL
corrections to the BFKL kernel, based on an explicit Monte Carlo integration
of the produced particles. We have demonstrated that in the standard
calculation of the NLL BFKL kernel, the NLL corrections arise from
Reggeon-Reggeon energies far higher than that which is probed at LL. This
exaggerates the apparent effect of the NLL corrections.

We have presented a method which will assign the NLL corrections to the
appropriate LL contributions. We are looking forward to reporting on the
gluon contribution to the NLL corrections, and on the implementation of the
new framework in the prediction for multi-jet events at the LHC.

\section*{Acknowledgements}
\label{sec:acknowledgements}
I would like to thank Robert S.~Thorne for discussions leading to the
localisation of a misprint in Ref.\cite{Fadin:1997hr}. I thank Einan Gardi,
Robert S.~Thorne and Bryan R.~Webber for many stimulating discussions.

\bibliographystyle{JHEP}
\bibliography{database}

\end{document}